\documentclass[apjl]{emulateapj}

\newcommand{\kev}{keV}
\newcommand{\fe}{Fe~K$\alpha$}
\newcommand{\etal}{et al.}
\newcommand{\fourc}{4C+74.26}
\newcommand{\mcg}{MCG--6-30-15}
\newcommand{\ergcms}{erg~cm$^{-2}$~s$^{-1}$}
\newcommand{\xmm}{\textit{XMM-Newton}}

\slugcomment{Accepted by ApJ Letters}

\shorttitle{Untruncated accretion disk in \fourc}
\shortauthors{Ballantyne \& Fabian}

\usepackage{times}
\usepackage{graphicx}

\begin{document}

\title{Evidence for an Untruncated Accretion Disk in the broad-line
  radio galaxy \fourc}


\author{D. R. Ballantyne\altaffilmark{1} and A. C. Fabian\altaffilmark{2}}
\altaffiltext{1}{Canadian Institute for Theoretical Astrophysics,
McLennan Labs, 60 St. George Street, Toronto, Ontario, Canada M5S 3H8;
ballantyne@cita.utoronto.ca} 
\altaffiltext{2}{Institute of Astronomy, Madingley Road, Cambridge,
  U.K. CB3 0HA; acf@ast.cam.ac.uk} 

\begin{abstract}
We present evidence for a broad, ionized \fe\ line in the \xmm\
spectrum of the broad-line radio galaxy (BLRG) \fourc. This is the
first indication that the innermost regions of the accretion flow in
BLRGs contain thin, radiatively efficient disks. Analysis of the 35~ks
\xmm\ observation finds a broad line with an inner radius close to the
innermost stable circular orbit for a maximally spinning black
hole. The outer radius of the relativistic line is also found to be
within 10~gravitational radii. The \fe\ line profile gives an
inclination angle of $\sim 40$\degr, consistent with the radio
limit. There are two narrow components to the \fe\ complex: one at
6.4~\kev\ from neutral Fe, and one at 6.2~\kev. These may form the
blue and red horns of a diskline from farther out on the disk, but a
longer observation is required to confirm this hypothesis. We discuss
the implications of this observation for models of jet production, and
suggest that BLRGs and radio-loud quasars will have larger than
average black hole masses, thus resulting in thicker accretion flows
close to the black hole.
\end{abstract}


\section{Introduction}
\label{sect:intro}
A long-standing problem in modern astrophysics is to understand the
production of large-scale, relativistic jets. In terms of the active
galactic nuclei (AGN) phenomenon, this problem has traditionally
manifested itself as understanding the underlying physical difference
between the majority radio-quiet population and the minority
radio-loud sources \citep[e.g.][]{kel89,wc95}. A potentially promising
way to elucidate this difference is to compare X-ray observations of
the two populations, since the X-rays originate from the inner regions
of the accretion flow, as do relativistic jets
\citep{bz77,bp82,koi04}. Studies of broad-line radio galaxies (BLRGs)
with \textit{ASCA}, \textit{RXTE} and \textit{BeppoSAX} seemed to
indicate that they had weaker reflection features and \fe\ lines then
their radio-quiet Seyfert~1 counterparts
\citep*{eh98,woz98,era00}. This could be explained if BLRGs contain a
truncated accretion flow, where the geometrically thin, radiatively
efficient accretion disk transforms close to the black hole into an
thicker, tenuous and radiatively inefficient configuration such as an
ADAF \citep[e.g.,][]{ny95}. This idea has some theoretical
justification as models of jet production have emphasized the
importance and connection of the polodial magnetic field to the inner
accretion flow \citep*{mei99,lop99,mei01,lpk03}. Furthermore, Galactic
black holes show radio emission (and, when resolved, jet structure)
only in the low-luminosity 'hard' or 'power-law dominated' state
\citep*[e.g.,][]{fbg04} where the accretion rate is expected to be
very low compared to the Eddington rate.

An alternative explanation for the weak reflection features observed
in BLRGs was put forward by \citet*{brf02}, who suggested an origin in
an ionized non-truncated accretion disk. This might be expected if the
accretion rate was a much larger fraction of Eddington. More
sensitive observations of BLRGs by \xmm\ were expected to discriminate
between the two explanations, but, until now, the results have been
inconclusive (NGC~6251, \citealt{gli04}; 3C~120, \citealt*{bfi04};
3C~111, \citealt{lew05}). In this Letter, we present the first
evidence, in the form of a broad \fe\ line, for an untruncated
accretion disk in a luminous BLRG. The line is broad enough to require
emission within the innermost stable circular orbit (ISCO) of a
Schwarzschild black hole, and, along with the reflection continuum, is
well fit by a moderately ionized reflector. Thus, this observation may
provide important constraints on the geometry of the inner accretion
flow in jet producing systems.

In the next section, we provide a brief introduction to the BLRG
\fourc, and then describe the \xmm\ observation and data reduction in
\S~\ref{sect:obs}. The spectral analysis is presented in
\S~\ref{sect:results}, and then we discuss the implications of the
results in the final section. Throughout this paper, a \textit{WMAP}
cosmology ($H_0=70$~km~s$^{-1}$~Mpc$^{-1}$, $\Omega_{\Lambda}=0.73$,
$\Omega_0=1$; \citealt{spe03}) is assumed.

\section{The Broad-line Radio Galaxy \fourc}
\label{sect:fourc}
\fourc\ ($z=0.104$; \citealt{ril88}) is a low luminosity radio-loud
quasar notable for its 10\arcmin\ radio lobes \citep{ril88}. A
one-sided jet has been detected by the VLA \citep{rw90} and at pc
scales by VLBI \citep{pear92}. The lack of a counter jet gives a limit
to the inclination angle of $i \la 49$\degr\ \citep{pear92}. The
total radio luminosity of the source places it on the border between
the FRI/II classes, although the observed structure is similar to a
FRII source \citep{ril88}. \citet{wu02} quote a bolometric luminosity of $L_{\mathrm{bol}} \approx 2\times
10^{46}$~ergs~s$^{-1}$. Optical spectra of \fourc\ show very broad
permitted lines, with \citet{cor97} measuring a H$\beta$ FWHM of
11,000~km~s$^{-1}$, although other authors give values closer to
8000~km~s$^{-1}$ \citep{ril88,brink98,rob99}. Using the \citet{cor97}
value of the FWHM and the \citet{kas00} radius-luminosity relation,
\citet{wu02} estimate the black hole mass in \fourc\ to be $\sim
4\times 10^9$~M$_{\odot}$.

A 23~ks \textit{ASCA} observation of \fourc\ yielded inconclusive
results despite three separate analyses.  \citet{brink98} and
\citet{rt00} uncovered a \fe\ line at the ~97\% level, but
\citet{sem99} claimed the line was significant at $>99$\%. The data
could not determine if the line was broadened. The spectrum showed
evidence for hardening at high energies, which \citet{sem99} fit by a
separate power-law component, attributed to jet emission.  When the
data was fit by reflection models, reflection fractions much greater
than unity were found ($R \sim 6$, \citealt{brink98}; or $R \sim 3$,
\citealt{rt00}). An unresolved line was clearly detected in the 100~ks
\textit{BeppoSAX} observation reported by \citet*{hse02}. The broadband
fits indicated a Compton reflection component at the 98.7\% level with
$R \sim 1$. This fact, along with the flat light
curve at high energies, shows that the X-rays are not significantly
contaminated by jet emission.

\section{Observations and Data Reduction}
\label{sect:obs}
\xmm\ \citep{jan01} observed \fourc\ during revolution 762 between
2004 February 6 13:57:42 and 2004 February 6 23:22:54. The European
Photon Imaging Camera (EPIC), comprising of two MOS \citep{tur01} and
one pn \citep{str01} detectors, was operated in large window mode with
the medium optical filter in place. Calibrated event lists were
extracted from the observation data files using the standard
processing chains (\textsc{epchain} and \textsc{emchain}) provided by
the \xmm\ Science Analysis System (SAS) v.6.1. A circular region with
radius 115\arcsec\ (as suggested by the SAS `extraction optimizer')
was employed to extract a pn spectrum of \fourc\ that included both
single and double events. The background spectrum was extracted from a
source free area on the same CCD using a circular region with a radius
of 60\arcsec. Contamination from a background flare $\sim 22$~ks into
the observation was removed with the use of a good-time interval file.
The SAS task \textsc{epatplot} confirmed that the pn spectrum was free
from pileup, but found evidence for it in the MOS spectra.  These data
were thus excluded from spectral analysis. The final, background
subtracted pn spectrum contains 252 050 counts obtained in 28.8~ks of
good exposure time, for a mean count rate of 8.6~s$^{-1}$. The
response matrix and ancillary response file for the pn spectrum were
generated using the \textsc{rmfgen} and \textsc{arfgen} tools within
the SAS.

\section{Spectral Analysis}
\label{sect:results}
As this paper is concentrating on the \fe\ region of the spectrum, we
restrict our analysis to the 2--12~\kev\ region of the pn data (full
broadband fits will be presented in a forthcoming paper). Prior to spectral
analysis, the data were grouped to have a minimum of 20 counts per
bin.  Model fitting was performed with XSPEC v.11.3.1p
\citep{arn96}. The uncertainties quoted on the best-fit parameters are
the 2$\sigma$ errorbars for one parameter of interest (i.e. $\Delta
\chi^2 = 2.71$). The Galactic absorption column of $1.19\times
10^{21}$~cm$^{-2}$ \citep{dl90} is included in all the spectral fits
presented below, and is modeled with the \textsc{tbabs} code
\citep*{wam00} within XSPEC.  All energies are quoted in the rest
frame, while the figures are plotted in the observed frame.

\begin{figure}
\centerline{
\includegraphics[angle=-90,width=0.45\textwidth]{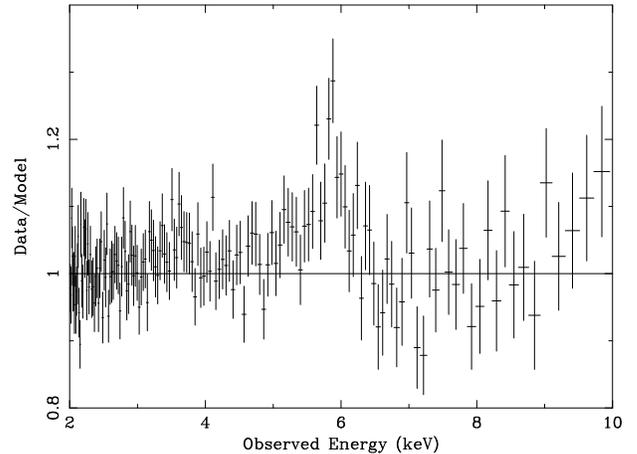}
}
\caption{Plot of the data-to-model ratio between 2 and 10~\kev\
  (observed frame) when an absorbed power-law model has been fit to
  the 2--12~\kev\ data, excluding the range between 5 and 7~\kev\
  (rest frame). The shape of the \fe\ line is evident in the
  residuals, and includes a narrow core and a weak red tail. Another
  possible narrow component is also apparent at $\sim 6.2$~\kev\ (rest
  frame). The figure also illustrates the hardening at high energies
  due to the reflection continuum.}
\label{fig:ratio}
\end{figure}

To get an overall sense of the \fe\ region in \fourc, we begin by
fitting the 2--12~\kev\ spectrum with a simple power-law model, but
excluding the region between 5 and 7~\kev. A good fit
($\chi^2/$d.o.f.$=857/939$; d.o.f. = degrees of freedom) was obtained
with $\Gamma=1.69\pm 0.02$. Figure~\ref{fig:ratio} plots the residuals
to this fit including the previously ignored energy range.
The figure reveals a well resolved \fe\ line profile consisting of a
broad red wing and potentially two narrow components contributing to
the line core. At higher energies, the residuals show the spectral
hardening indicative of a reflection continuum. 

\begin{deluxetable*}{cccccccccc}
%\tabletypesize{\small}
%\rotate
\tablewidth{0pt}
\tablecaption{\label{table:fits}Model parameters from fitting the
  2--12~keV (observed frame) spectrum of \fourc.}
\tablehead{
\colhead{Model} & \colhead{$\Gamma$} & \colhead{$E_{\mathrm{G}}$} &
\colhead{EW} & \colhead{$R$} & \colhead{$r_{\mathrm{out}}$} &
\colhead{$i$} & \colhead{EW$_{\mathrm{L}}$} & \colhead{$\log \xi$} & \colhead{$\chi^2$/d.o.f.}}
\startdata
PL+G\tablenotemark{a} & 1.71$\pm 0.02$ & 6.46$^{+0.04}_{-0.02}$ & 45$^{+9}_{-15}$ &
\nodata & \nodata & \nodata & \nodata & \nodata & 1163/1301 \\[1mm]
PL+G\tablenotemark{a}+L\tablenotemark{b} & 1.74$\pm 0.02$ & 6.46$^{+0.04}_{-0.03}$ & 28$^{+11}_{-12}$ & \nodata &
7.1$^{+8.0}_{-2.4}$ & 45$^{+11}_{-14}$ & 238$^{+77}_{-80}$ & \nodata &
1138/1298 \\[1mm]
PEXRAV\tablenotemark{c}+G\tablenotemark{a}+L\tablenotemark{b} & 1.81$^{+0.06}_{-0.05}$ & 6.46$^{+0.05}_{-0.04}$ &
25$^{+11}_{-12}$ & 1.2$^{+0.7}_{-0.6}$ & 5.4$^{+3.2}_{-1.4}$ & 45\tablenotemark{f}
& 167$^{+69}_{-66}$ & \nodata &  1128/1298\\[1mm]
IONDISK*blr\tablenotemark{d}+IONDISK\tablenotemark{e} & 1.82$\pm 0.02$ & \nodata & 35 & 1\tablenotemark{f} &
6.3$^{+3.2}_{-3.0}$ & 37$^{+34}_{-6}$ & 154 &
2.64$^{+0.06}_{-0.08}$ & 1130/1299\\[1mm]
IONDISK*blr\tablenotemark{d}+IONDISK\tablenotemark{e}+G\tablenotemark{a}
& 1.82$\pm 0.02$ & 6.23$^{+0.04}_{-0.05}$ & 38 \& 20 & 1\tablenotemark{f} &
6.3$^{+3.5}_{-2.5}$ & 34$^{+10}_{-5}$ & 134 &
2.60$^{+0.07}_{-0.08}$ & 1121/1297 \\
\enddata
\tablenotetext{a}{$\sigma=0$\tablenotemark{f}}
\tablenotetext{b}{$r_{\mathrm{in}}=1.235$~$r_g$\tablenotemark{f},
  emissivity$=-3$\tablenotemark{f}, $E=6.4$~\kev \tablenotemark{f}}
\tablenotetext{c}{$E_{\mathrm{fold}}=200$~\kev \tablenotemark{f},
  abundances=solar\tablenotemark{f}}
\tablenotetext{d}{$r_{\mathrm{in}}=1.235$\tablenotemark{f},
  emissivity$=-3$\tablenotemark{f}}
\tablenotetext{e}{Reflection dominated, $\log \xi=1$\tablenotemark{f}}
\tablenotetext{f}{Parameter fixed at value}
\tablecomments{In the model descriptions, PL=power-law, G=Gaussian
  emission line, L=\citet{lao91} relativistic line, PEXRAV=neutral
  reflection continuum of \citet{mz95}, IONDISK=ionized
  reflection spectrum of \citet{rfy99}, and 'blr'=blurred with the
  Laor relativistic kernel. $E_{\mathrm{G}}$ is the energy of the
  added Gaussian line in \kev, EW is the equivalent width(s) of any
  narrow component in eV, $R$ is the reflection fraction,
  $r_{\mathrm{out}}$ is the outer radius of the relativistic emission
  line in $r_g$, $i$ is the inclination angle in degrees,
  EW$_{\mathrm{L}}$ is the equivalent width of the relativistic line
  in eV, and $\xi$ is the ionization parameter in the IONDISK model.}
\end{deluxetable*}
Results of the spectral fitting are show in
Table~\ref{table:fits}. All the fits provide a very acceptable
statistical fit to the data, but significant improvements were found
with the addition of a broad \fe\ component, reflection, and a narrow
line at $\sim 6.2$~\kev. After many trials with the \citet{lao91} and
'diskline' \citep{fab89} relativistic line models, the best fit was
found with an inner radius close to the ISCO for a spinning Kerr black
hole, and an outer radius at $\sim 6$~$r_g$, where $r_g=GM/c^2$ is the
gravitational radius for a black hole with mass $M$. Broad lines with
larger inner radii not only had larger $\chi^2$s ($\Delta \chi^2
\approx +10$), but gave inclination angles $i \sim 20$\degr, implying
that the physical size of the \fourc\ radio source is $\sim 3$~Mpc,
larger than almost any other known giant radio galaxy (GRG;
\citealt{lar01,lar04}). The broad Laor line results in a larger
inclination angle, and therefore a smaller physical size of $\sim
2$~Mpc, more in line with other GRGs. Allowing the line emissivity $\beta$
to vary did not improve the fit ($\beta=-3^{+4}_{-1.6}$).   

\begin{figure}
\centerline{
\includegraphics[angle=-90,width=0.45\textwidth]{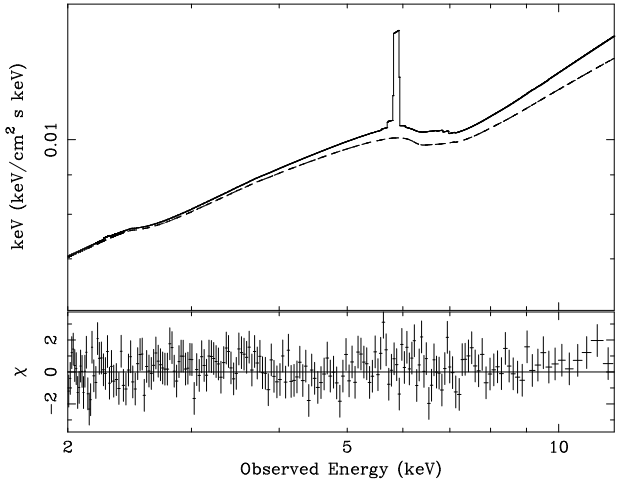}
}
\caption{The top panel shows the IONDISK*blr+IONDISK model from
  Table~\ref{table:fits}, with the solid line denoting the total model
  and the dashed line plots the blurred, ionized reflector. The
  unblurred, neutral reflector which accounts for the narrow 6.4~\kev\
  line is off the bottom of the plot. The lower panel plots the
  residuals (in units of standard deviations) when this model is fit to
  the \fourc\ data.}
\label{fig:model}
\end{figure}

The rest energy of the Laor line was fixed at the emission energy for
neutral iron at 6.4~\kev, but, when that line and the PEXRAV continuum
was replaced with the ionized reflection model of \citet*{rfy99}, the
ionization parameter was found to be tightly constrained at $\xi
\approx 400$ and predicted emission from helium like Fe at
6.7~\kev. For this model, denoted IONDISK*blr+IONDISK in
Table~\ref{table:fits}, a reflection dominated neutral continuum
accounted for the sharp core of the line. Figure~\ref{fig:model} plots
this model and the residuals to this fit.
The residuals show a possible additional emission feature associated
with the \fe\ complex. A narrow Gaussian added to the model was found
to be significant at the 99\% level, and had a best fit energy of
$6.2$~keV and EW of 20~eV. The origin of this line is unknown. It is
possible that it and the 6.4~\kev\ line may comprose the red and blue
horns of a diskline from farther out on the disk
\citep*[c.f.][]{tur02,yaq03,tkr04}. To test this, we replaced the
Gaussian in the PEXRAV+G+L model with a diskline with rest energy
6.4~\kev, an outer radius of 1000~$r_g$, and an emissivity of $-2$ (a
flatter emissivity is expected if the disk is warped at larger
radii). A good fit was obtained with $\chi^2/$d.o.f.$=1126/1299$, but
the inner radius was unconstrained ($r_{\mathrm{in}} > 52$~$r_g$).

The last model in Table~\ref{table:fits} yields a 2--10~\kev\ flux of
$F_{\mathrm{2-10\ keV}} = 2.43\times 10^{-11}$~\ergcms. The earlier
\textit{ASCA} and \textit{BeppoSAX} observations found
$F_{\mathrm{2-10\ keV}} = 1.7\times 10^{-11}$~\ergcms\ and $1.4\times
10^{-11}$~\ergcms, respectively, indicating that this \xmm\
observation caught \fourc\ in a higher flux state. The unabsorbed rest
frame 2--10~\kev\ \xmm\ luminosity is $L_{\mathrm{2-10\ keV}} =
6.6\times 10^{44}$~erg~s$^{-1}$. The total unabsorbed 0.3--12~\kev\
luminosity of \fourc\ is $L_{\mathrm{0.3-12\ keV}}=1.6\times
10^{45}$~erg~s$^{-1}$. Assuming this comprises about 10\% of the
bolometric luminosity, $L_{\mathrm{bol}}\approx 1.6\times
10^{46}$~erg~s$^{-1}$, in good agreement with the $L_{\mathrm{bol}}$
found by \citet{wu02}. A black hole mass of $4\times 10^9$~M$_{\odot}$
\citep{wu02} then gives an observed Eddington ratio of $\sim 0.04$,
consistent with the presence of a untruncated thin accretion disk over
an inner ADAF-like flow (which requires
$\dot{M}/\dot{M}_{\mathrm{Edd}} \la \alpha^2$, where $\alpha \sim 0.1$
is the \citet{ss73} viscosity parameter; \citealt{ree82}).

\section{Discussion}
\label{sect:discuss}
In the previous section we presented evidence that the X-ray spectrum
of \fourc\ exhibits a broad ionized \fe\ line extending very close to
a spinning black hole. This evidence is by no means conclusive (e.g.,
it is difficult to rule out the possibility that the line shape is
actually due to a complex series of absorbers), but it is
suggestive. A longer \xmm\ observation is required to confirm its
presence and extent. Below, we assume the line is as measured above,
and discuss how it may shed light on the problem of jet formation.

It may be constructive to compare the properties of \fourc\ to those
of the well known radio-quiet Seyfert~1 \mcg\ which also has a broad
line that implies a spinning black hole \citep{fab02}. What quality
(or qualities) allow \fourc\ to produce a strong radio jet and
inhibits one in \mcg?  Evidently, the spin of the black hole seems not
to be important as there is increasing evidence (albeit
circumstantial) that most black holes are spinning. This is based on
comparisons of the accreted black hole mass density (as judged from
the X-ray background) to the local density of black holes
\citep*{erz02,bar04}, and from simulations of black hole growth
including both mergers and accretion \citep{shap04}. Moreover, the
observation that radio emission is quenched as Galactic black holes
move from the low state to the high state precludes a strict spin
dependence. The accretion rates may also be ruled out as the governing
parameter since radio-loud quasars produce jets at high accretion
rates, as does the BLRG 3C~120 \citep{bfi04}.

Interestingly, the mass of the black hole in \fourc\ is estimated to
be $\sim 10^3\times$ larger than the one in \mcg\ \citep{bz03}. In
the last few years, other authors have presented evidence that
radio-loud AGN preferentially have large black hole masses
\citep{lao00,md02,mj04}, although there have been dissenting views
\citep{ho02,wu02b}. Assuming the same accretion rate and a radiation
pressure dominated inner disk, a larger black hole mass will increase
the scale height $H$ and decrease the density of the accretion flow
\citep{ss73}. The larger value of $H/r$ (where $r$ is the radius along
the disk) may enhance the polodial magnetic field over the lower mass AGN and
increase the chances of jet emission.

There now seems to be two types of sources that produce strong radio
emission \citep*{mcf04}. The first are low accretion rate objects, such as LINERs and
other low-luminosity AGN, and Galactic black holes in the low,
power-law dominated state \citep[e.g.][]{ho02}. These are the sources
that populate the fundamental plane of black hole activity recently
discovered by \citet*{mhd03}. The second class of sources have much
higher luminosities and accretion rates, and we would argue that it includes BLRGs
and radio-loud quasars. These objects would have
untruncated, radiatively efficient accretion disks, but relatively
high black-hole masses and thus a larger $H/r$ close to the black hole
than radio-quiet Seyfert~1s\footnote{\fourc\ has a relatively low
$\xi$ (Table~\ref{table:fits}), suggesting that the disk must be
fairly dense or not highly illuminated. This may be understood if, as
is often suggested \citep[e.g.][]{br02}, the ionization parameter is
correlated with the accretion rate. This would also explain the
possible very
high $\xi$ inferred from 3C~120 \citep{bfi04}.}. Those sources which
fall along the fundamental plane are powered by radiatively
inefficient accretion flows, which are also thick and would have
enhanced polodial magnetic fields. The connection between all jet
emitting sources would then be the structure and magnetic strength of
the inner accretion flow. Hopefully, this hypothesis can be confirmed
by numerical simulations of jet formation.

In summary, the broad \fe\ line detected in the \xmm\ observation of
\fourc\ rules out the possibility of a truncated accretion disk in
BLRGs. Rather, the observation strengthens the scenario that the disk
thickness (in the sense of $H/r$) close to the black hole is the
important parameter for jet production. For high accretion rate
sources, such as BLRGs, the thickness is a result of a relatively high
black hole mass. In low accretion rate sources, such as low-state
X-ray binaries, it is caused by a radiatively inefficient accretion
flow.

\acknowledgments

This research is based on observations obtained with \xmm, an ESA
science mission with instruments and contributions directly funded by
ESA Member States and NASA. DRB acknowledges financial support by the
Natural Sciences and Engineering Research Council of Canada. ACF
thanks the Royal Society for support.

\end{document}